\documentclass{IEEEcsmag}

\usepackage[colorlinks,urlcolor=blue,linkcolor=blue,citecolor=blue]{hyperref}

\usepackage{upmath}
\usepackage{caption} 
\usepackage[utf8]{inputenc}
\usepackage{wrapfig}

\jvol{}
\jnum{}
\paper{}
\jmonth{}
\jname{}
\pubyear{}

\begin{document}

\title{Using a Cyber Digital Twin for Continuous Automotive Security Requirements Verification}

\author{Ana Cristina Franco da Silva}
\affil{University of Stuttgart}

\author{Stefan Wagner}
\affil{University of Stuttgart}

\author{Eddie Lazebnik}
\affil{Cybellum}

\author{Eyal Traitel}
\affil{Cybellum}

\markboth{Department Head}{Using a Cyber Digital Twin for Continuous Automotive Security Requirements Verification}

\begin{abstract}%
A Digital Twin (DT) is a digital representation of a physical object used to simulate it before it is built or to predict failures after the object is deployed. 
In this article, we introduce our approach, which applies the concept of a Cyber Digital Twin (CDT) to automotive software for the purpose of security analysis. In our approach, automotive firmware is transformed into a CDT, which contains automatically extracted, security-relevant information from the firmware. 
Based on the CDT, we evaluate security requirements through automated analysis and requirements verification using policy enforcement checks and vulnerabilities detection. 
The evaluation of a CDT is conducted continuously integrating new checks derived from new security requirements and from newly disclosed vulnerabilities. We applied our approach to about 100 automotive firmwares. In average, about 600 publicly disclosed vulnerabilities and 80 unknown weaknesses were detected per firmware in the pre-production phase.
Therefore, the use of a CDT enables efficient continuous verification of security requirements.
\end{abstract}

\maketitle

\chapterinitial{A common challenge} in verifying security requirements is that they are often abstract and, thereby, underspecified~\cite{Fernandez2017}. Examples of such security requirements in the automotive domain are those provided by the UNECE WP.29 regulation~\cite{UniNations2020} and the ISO/SAE 21434 standard~\cite{ISO-21434-2020}. They mostly prescribe processes and methods but do not specify how security can be achieved in a concrete system. This leads to various problems in the development but also in the quality assurance of these systems. Cruzes et al.~\cite{Cruzes2017} argue that testing non-functional requirements, such as security, is a great challenge due to cross-functional aspects of testing and the lack of clarity of their needs. 
Concrete requirements help us to better check the security of our systems, however, we also need solutions that are able to cope with abstract requirements as well, such as those within the UNECE WP.29 regulation and the ISO/SAE 21434 standard.

A further challenge in verifying security requirements results from outsourcing practices and use of third-party software in the automotive industry~\cite{Ciravegna2013}. 
Due to theses practices, the complete source code is usually not available, but rather only the binaries are available for security analyses. 
In this case, techniques that do not require source code, such as black-box techniques~\cite{Liu2012}, need to be used in verifying security requirements of automotive software. However, to efficiently identify vulnerabilities in the firmware, relevant information, such as hardware and software configurations, used software packages, operating system (OS) and application-specific settings, need to be extracted from these binaries.

In this article, we present our approach to verify security requirements by tackling the aforementioned challenges. Our approach is based on the automated creation of a \emph{Cyber Digital Twin~(CDT)} for an automotive firmware, which is used to conduct continuous security vulnerability analyses and policy enforcement checks. Found vulnerabilities and policy violations are used to identify which security requirements have not been satisfied by the analyzed firmware.
The Digital Twin (DT) is a concept to digitally represent physical objects to simulate them before they are built or to enable predictive maintenance~\cite{Boschert2016}. 
The CDT transfers the concept of the DT to software as well~\cite{CTL2020}~(cf. Section 1). 
In this article, a CDT is a digital representation of an automotive firmware, which is a class of software stored and executed on Electronic Control Units~(ECUs). 
Such CDTs provide many advantages with respect to automated security evaluation of automotive firmwares, especially considering that large parts of software are built based on other software libraries and frameworks, which are not available as source code. In the case of automotive Original Equipment Manufacturers (OEMs), this even means most of the software in the ECUs in the cars they build have software on them that they can only evaluate as black boxes. 

\begin{table}\normalsize
	\centering
	\renewcommand{\arraystretch}{1.5}
		\begin{tabular}[t] {| p{7cm} |}
			\hline
\textbf{Sidebar 1: Black-Box Vulnerability Analysis}\\
There are alternatives to the CDT approach to perform black-box analyses for vulnerabilities in software. The main approaches to vulnerability analysis are static analysis of the source code or black-box testing often using fuzzing techniques~\cite{Liu2012}. Static analysis usually requires access to the source code of the software and, hence, cannot be used in a black-box way.\\
In case of black-box testing, there is a common focus, for example, on crawling and then testing web applications for side-channel vulnerabilities~\cite{Chapman2011} or SQL injection vulnerabilities~\cite{Djuric2013}. Yet, these approaches need an executable system often installed in a real or realistic environment.\\
Regarding penetration tests, we do not primarily evaluate security requirements by simulating attacks to the firmware, but rather by conducting different analyses on the CDT of the firmware based on pattern matching techniques. However, once the analyses detect that the firmware uses a software library affected by a known vulnerability in Common Vulnerabilities and Exposures (CVE) databases, the CDT is executed in a sandbox and penetration tests are carried out to trigger the vulnerability as described in the CVE record.\\
			\hline
		\end{tabular}
			\label{tab:sidebar1}
\end{table}

Our approach automatically extracts important information from firmware binaries, such as bill of materials, interfaces, control and data flow, which are needed for security analysis activities. 
This extracted information, which constitutes the CDT, enables us to discover security vulnerabilities in the firmware, and furthermore, to check if certain security requirements of the firmware are not fulfilled. 
In this context, a \emph{vulnerability} (\url{https://cve.mitre.org/about/terminology.html\#vulnerability}) is defined as a weak spot in the automotive firmware, which can be exploited by intruders and negatively affect the confidentiality, integrity or availability of the firmware. We discuss alternative approaches to our CDT approach in Sidebar~1.

The CDT approach has been applied to hundreds of automotive firmware to detect known vulnerabilities in the used libraries, which are publicly disclosed in Common Vulnerabilities and Exposures (CVE) databases. Furthermore, it has also been applied to discover unknown weaknesses, such as \textit{heap overflow} and \textit{use after free}, based on the classification provided by the Common Weakness Enumeration (CWE).
We describe our approach including the analysis method and how it can be used to verify security requirements in the following.

\section{1. DIGITAL TWIN AND CYBER DIGITAL TWIN}\label{sec:background}
The DT concept comprises not only simulation but also monitoring of existing physical objects including their current states and processes, so that DTs are always in-sync with their physical counterparts to enable reliable monitoring~\cite{Grieves2017}. 
The concept of using twins has emerged within NASA's Apollo program and is still in use today, where space vehicle twins are used for training and simulation purposes~\cite{Glaessgen2012}. 
The DT concept has been applied to manufacturing, introduced by Michael Grieves in 2002, and has gained attention in further areas such as automotive and IoT. An exemplary software solution in manufacturing is the GE Digital Predix Platform (\url{https://www.ge.com/digital/iiot-platform}), which is an IIoT platform for creating and tracking DTs of industrial systems. %

The CDT extends the DT concept to also represent software objects that can be regarded only as black-boxes. We define a CDT as a digital representation of an automotive firmware, which only consists of binaries.
This digital representation is enhanced with critical information for security analyses in both pre-production and post-production phases. Hence, the CDT enables vulnerability management and security requirements evaluation of a firmware throughout its lifecycle.

The use of DTs for software has also been proposed in other works to identify cybersecurity issues. Hadar et al.~\cite{Hadar2020} use the term Cyber Digital Twin Simulator, in which a CDT is created based on active networks and attack graph analytics. This CDT automatically collects information about the computer network infrastructure, associates it with attack tactics, measures the efficiency of implemented security controls requirements and automatically detects missing ones. The resulting CDT corresponds to a knowledge graph inferencing model of the active network. 

\section{2. THE CYBER DIGITAL TWIN APPROACH}
\begin{figure*}[t!]
\centerline{\includegraphics[width=0.9\textwidth]{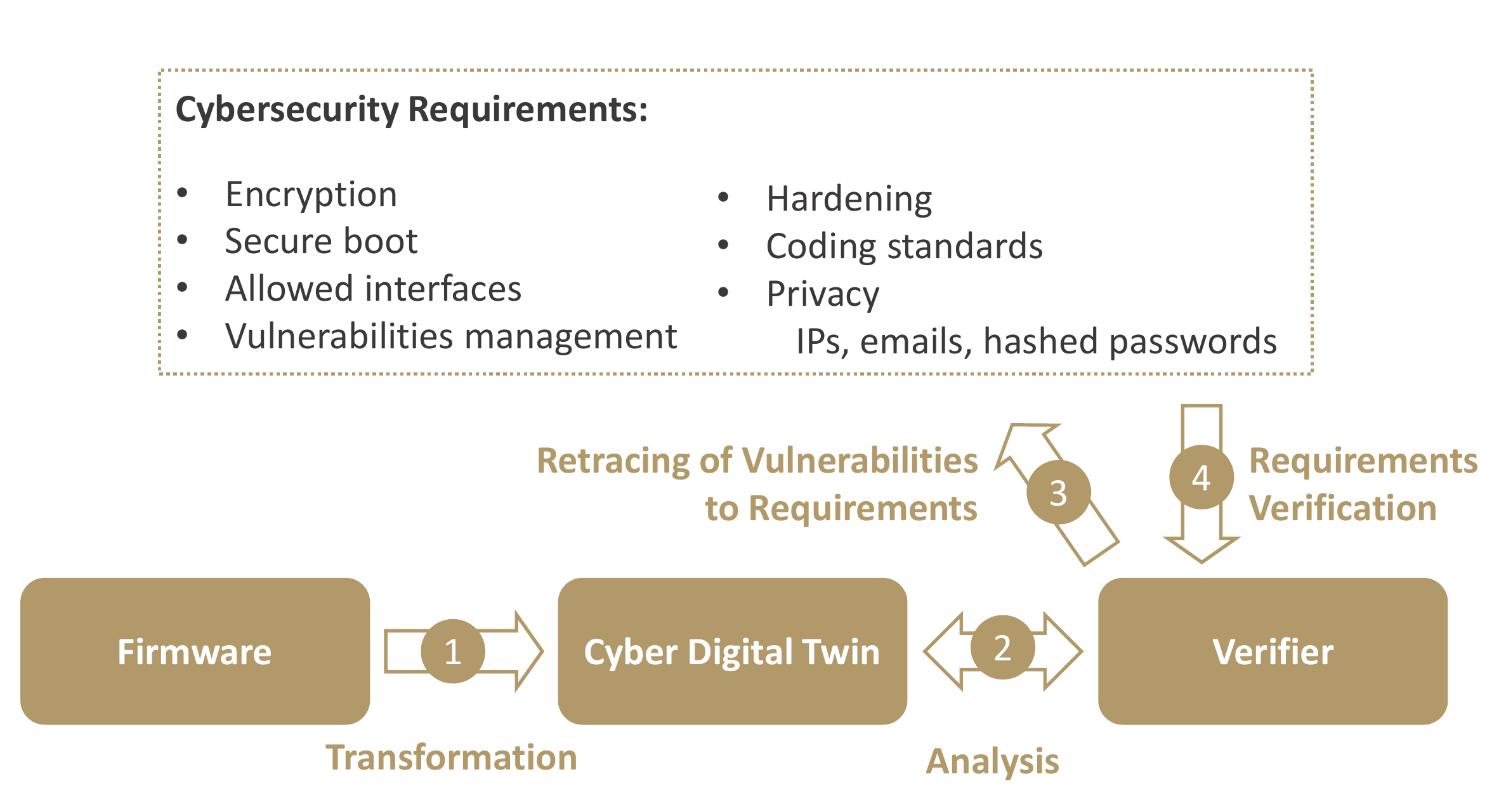}}
\caption{Using Cyber Digital Twins for continuous automotive security requirements verification}
\label{fig:approach}
\end{figure*}

In this section, we present our CDT approach, which is depicted in Figure~\ref{fig:approach}.

\subsection{2.1. Transformation of a Firmware into a CDT}

As first step, an automotive firmware binary, such as an infotainment system, is automatically transformed into a corresponding CDT. %

During the creation of the CDT, all available interfaces, all used software libraries and further information are discovered automatically and added to the model underlying the CDT.
If available, software inventories of suppliers are also used. 
Not only interfaces and employed software libraries are automatically extracted, but a much more comprehensive set of information is extracted and added to the CDT as summarized in Table~\ref{tab:TheCyberDigitalTwinModel}.
Possible interfaces are for example \textit{Bluetooth}, \textit{Controller Area Network (CAN bus)} and \textit{GPS}. Examples of normally used software libraries in automotive firmwares are the \textit{openssl} and \textit{sqllite} libraries.
In addition, the CDT also includes the interactions among all items in Table~\ref{tab:TheCyberDigitalTwinModel}, and any additional configuration information and metadata describing the firmware and its behavior.
At the core of the CDT – as in a normal Digital Twin – is the software bill of materials (SBoM) that lists the software components of the firmware. 

The creation of the CDT consists of: the \emph{firmware extraction}, the \emph{software composition analysis (SCA)} and \emph{the licensing analysis}. 
Automotive firmwares are available in many different formats, such as flash device dumps, file system images, virtual machine images, container images, compressed files, or, in most cases, a hierarchical combination of different formats.
The \emph{firmware extraction} process identifies the correct format of each file, recursively iterates through the file tree and extracts information using the suitable extraction code. This extraction code does not know if the result of a specific file’s extraction is correct, therefore this process requires validation steps to ensure that the result resembles the file tree structure and contents of the original firmware.

In the \emph{SCA}, the extracted files are scanned thoroughly to detect unique software components. This is performed by pre-processing executable objects, cleaning up irrelevant file sections, then comparing contents and attributes of the files with a large database of mapped components. Software components can be uniquely identified through certain file paths, file names, or unique strings. Others can be detected using direct parsing of configuration data of installed software within the firmware, for example, package management configuration databases in the firmware. 
The result of this analysis is the SBoM.%

In the \emph{licensing analysis}, the SBoM is analyzed to determine the software licenses, copyrights and legal notices related to each software component. This is important for licensing practices such as the ones defined by the OpenChain standard (\url{https://www.openchainproject.org}).

\begin{table*}\normalsize
	\centering
	\renewcommand{\arraystretch}{1.5}
		\begin{tabular}[t] {| p{5cm} | p{11cm} |}
			\hline
				\textbf{Hardware bill of materials (HW BoM)} & The detected hardware components and hardware configuration: (i) CPU architecture, (ii) CPU bit size, (iii) video card, (iv) network interfaces, (v) other peripheral components. This configuration can be derived from static analysis of configuration files, drivers and software libraries detected from the firmware.\\ \hline
				\textbf{Network interfaces} & The detected network interfaces, including physical communication interfaces such as Ethernet, USB, Wifi, Bluetooth, CAN bus, cellular (3G/4G/5G), radio and Zigbee, as well as logical communication protocols such as SMS.\\ \hline
				\textbf{Software bill of materials (SBoM)} & The detected list of distinct software components including libraries names and used versions. These components can be a mixture of open-source software (OSS), third-party commercial software, or first-part developed software libraries or applications. \\ \hline
				\textbf{Operating system} & The operating system - open-source (Linux, Android) or commercial (QNX, VxWorks etc.). \\ \hline
				\textbf{Operating system settings} & Operating system configurations and settings. For example, the Linux /etc files.\\ \hline
				\textbf{Kernel configuration} & The detected kernel configuration - the list of enabled modules and settings (such as sysctl parameters in Linux).\\ \hline
				\textbf{OS-level security configuration} & Specific operating system security configurations, for authentication or other security mechanisms.\\ \hline
				\textbf{Memory management and mapping} & The operating system memory management system, its configuration and actual usage pattern by the executables within the firmware.\\ \hline
				\textbf{User credentials} & All the user authentication details - user names, passwords, tokens etc.\\ \hline
				\textbf{Firewall configuration} & The built-in firewall configuration and rules.\\ \hline
				\textbf{Application frameworks} & Application frameworks in use in the firmware and their configuration. For example - some commercial frameworks exist which provide additional application-level support. \\ \hline
				\textbf{APIs} & The API sets that are available to use and the actual ones in use.\\ \hline
				\textbf{Application configuration} & Application-specific configurations and settings. For practicality purposes, the focus is mostly on settings that have an impact on security.\\ \hline
				\textbf{Encryption mechanisms and flows} & All encryption services and communication protocols using encryption, and their use by different systems and applications in the firmware.\\ \hline
				\textbf{Encryption keys} & Public and private encryption keys that are used to access internal and external services.\\ \hline
				\textbf{Control and data flow graph representation} & The actual applications code - the control and data flows.\\
			\hline
		\end{tabular}
	\caption{The Cyber Digital Twin model}
	\label{tab:TheCyberDigitalTwinModel}
\end{table*}

\subsection{2.2. Analysis of the Cyber Digital Twin}

In this step, the resulting CDT of a firmware is first analyzed in the pre-production phase to discover threats, security vulnerabilities, and security policy violations. 
In this way, OEMs can solve issues before the firmware is released.

Based on the hardware and software bill of materials, the CDT is analyzed using pattern recognition techniques, to find potential vulnerabilities, which are published in Common Vulnerabilities and Exposures (CVE) databases. 
For each software library in the SBoM, all related CVEs are automatically identified by querying the CVE database through APIs. 

Each CVE is tagged by the numbering authorities to specific software components according to the Common Platform Enumeration (CPE, \url{https://nvd.nist.gov/products/cpe}). The CPEs that make up a firmware result from the SCA in the CDT creation. This matching process between the list of software components (CPEs) and vulnerabilities (CVEs) is performed by querying the vulnerability databases.
However, this matching is not enough, since it creates a CVE list which is very generic. Many of the CVEs will not be relevant for the analyzed firmware. The existence of a certain software component in a specific version does not imply that all the vulnerabilities listed for that component are relevant. 
Therefore, the vulnerability analysis engine also considers context information of the firmware. 
The context information of the component itself is analyzed from the firmware, e.g., from its operating system, specific OS and kernel configurations, or the file’s CPU instruction set. In addition, security analysts provide additional input on additional context information that cannot be detected from the firmware itself.
The context information is analyzed from metadata available in the vulnerability database, from the vulnerability description, and from further information about the vulnerability available on the Internet. This analysis is performed using various text analysis techniques.

The CDT approach also performs a \emph{binary code analysis} to find unknown vulnerabilities. This analysis consists of many phases. Some of these run concurrently with others, and some rely on information from others. They are in serial order:

\begin{enumerate}
\item Binary File Loading: The proprietary files are loaded one at a time to the unknown detection engine, and the appropriate CPU architecture is detected.
\item Section Mapping: Mapping the data and code sections.
\item Instruction Set Disassembler: The assembly code is disassembled and mapped to an Intermediate Language (IL). This enables the analysis engine to support a variety of CPU instruction sets independently from the vulnerability analysis itself, which is performed on the IL.
\item Function Reconstruction: The functions are reconstructed from the instructions stream.
\item Function Parameters and Stack Analysis: Detection of function parameters and the stack usage.
\item Control Flow Graph (CFG): Building the intra-function control flow, and direct inter-function control flow transitions.
\item Data flow analysis: Analyzing the data flow in and out of functions. This phase includes variable tracing and taint checking.
\item Remediation Analysis: Analysis and identification of possible remediations to detected vulnerabilities.
\item Dynamic Execution: Dynamic execution of the models for the purpose of determining the validity of a given vulnerability.
\end{enumerate}

After the firmware has been released, the corresponding CDT is maintained and continuously monitored to detect and evaluate changes: in the firmware (e.g., through updates), in the CVE databases (e.g., through newly disclosed vulnerabilities), and in the security requirements (e.g., through introduction of new security policy regulations). That is, upon the mentioned changes, further analyses are necessary to be conducted.
For example, in case of a newly disclosed CVE in a used software library, the CDT is re-analyzed as described above to check if the firmware is affected by it. 

The CDT enables vulnerability management of an automotive firmware throughout its lifecycle. The continuous monitoring of the CDT is highly helpful and necessary since new vulnerabilities are discovered every day, and furthermore, already in production firmware might need to be updated a few times a year, what might in turn introduce further vulnerabilities. Consequently, the results of the analysis in pre-production will become deprecated over time. 

\subsection{2.3. Requirements Verification}
From experiences of working and conducting research on security solutions for the automotive industry, we identified several common security requirements. Examples for such requirements are the employment of \textit{data encryption}, \textit{secure boot} and \textit{system hardening techniques}. Furthermore, \textit{coding standards}, such as from AUTOSAR, and data privacy measures, such as \textit{hashed passwords}, should also be used. The\textit{ management of vulnerabilities} throughout the whole firmware lifecycle is a common requirement as well, since new vulnerabilities are discovered from time to time and consequently databases of publicly known cybersecurity vulnerabilities are continuously expanded.
The fulfillment of such requirements as well as the compliance to security standards enables higher levels of security in the firmware. Consequently, the verification and evaluation of security requirements of the firmware is a crucial task. 

In this context, cybersecurity requirements for an automotive firmware are specified from security policies, such as those in UNECE WP.29 and ISO/SAE 21434, or from discussions with stakeholders.
To check if the analyzed firmware meets its specified cybersecurity requirements, we have designed the component \emph{Verifier} (cf.~Figure~\ref{fig:approach}), which receives as input the CDT analysis results as CSV report files. Such a report includes a comprehensive list of known vulnerabilities with corresponding CVE IDs and CWE IDs and unknown weaknesses with corresponding CWE IDs.
We use algorithms to retrace the discovered vulnerabilities based on their CWE IDs to cybersecurity requirements. That is, we check if a specific vulnerability can be traced back to one or more requirements, and thereby identify the non-fulfillment of the retraced requirements. %
For this, mappings of CWEs to the specified requirements for the firmware need to be created. In this work, we have used mappings that were created manually for a subset of requirements based on the UNECE WP.29 regulation. 

For example for the software library \emph{sqlite v3.31.1}, the CVE-2020-11656 was published in the NIST national vulnerability database (NVD). This CVE reports referencing memory after it has been freed and is classified under the CWE-416 (use after free). 
This issue might enable the execution of arbitrary code that can be exploited by attackers. Based on the classification provided by the CWE of the CVE, this kind of vulnerability can be traced back to system hardening requirements (mapping “\textit{CWE-416: use after free $\leftrightarrow$ system hardening requirement}”). 
That is, an automotive firmware using the mentioned library and version does not fulfill a system hardening requirement, in case such a requirement was specified for the firmware.

Abstract requirements are still challenging, however. Yet, through the mapping of them to concrete CVEs and CWEs, such requirements are enriched with valuable information that can be leveraged within requirement verification and also for the creation or update of further test cases.   
As in the analysis of the CDT described in Section 2.2, the requirement verification is also conducted continuously as response to changes in the firmware, in the CVE databases or in the specified requirements, which results in the generation of new reports. 

In future work, we plan to automate the creation of the mappings based on established techniques, such as keyword extraction in Natural Language Processing (NLP), to process and better match CWE descriptions to requirements provided in cybersecurity regulations or in software requirements specifications (SRS).

In summary, the component \textit{Verifier} evaluates security requirements based on the SBoM, their identified vulnerabilities, and mappings from corresponding CWEs to specified requirements. 
For example, it checks if the firmware uses hardening techniques helping to avoid security attacks, such as address space layout randomization (ASLR). 
The mappings of CWEs and requirements are created and reviewed manually by security experts. Currently, we are designing a semi-automated solution to reduce the manual effort, in which such mappings are created automatically and reviewed by security experts to correct possible mismatches.

\section{3. RESULTS}

To comply with cybersecurity regulations, OEMs need to demonstrate that they use a cybersecurity management system capable of ensuring adequate security of vehicles~\cite{UniNations2020}. The CDT approach supports OEMs thereby through security analysis and vulnerabilities management in the development and post-production phase. 

To explain how the CDT approach is used to continuously verify requirements, we walk through the steps taken to verify a cybersecurity requirement, which was derived from the UNECE WP.29 cybersecurity regulation (4.3.6 \emph{Threats to vehicle data/code}, in Annex A, Table A1: List of vulnerability or attack method related to the threats)~\cite{UniNations2020}. In the exemplary requirement, firmwares shall prevent unauthorized access to privacy information. More concretely, to avoid privacy information leakage, adequate security technologies and mechanisms should be employed that validate and prevent privacy information leakage to happen. Examples of privacy information are personal identity, payment account information, passwords, e-mail addresses, or IP addresses.

First, the CDT is created for the firmware in question. Second, a security analysis is conducted, in which known vulnerabilities are searched for the libraries in the software bill of materials. In this analysis, the CDT in general is also checked for unknown vulnerabilities and security policies violations, such as system hardening misconfigurations. 
Afterwards, each discovered vulnerability is traced back to corresponding requirements based on its threats, vulnerabilities and attack methods. If no vulnerabilities could be traced back to the aforementioned exemplary requirement, this requirement will be verified as fulfilled. 
If at least one vulnerability can be traced back to the requirement, this requirement will not be fulfilled. Finally, the security analysis results, e.g., the uncovered vulnerabilities of the firmware in question, are described in a comprehensive report, which is generated automatically.  Note that security analyses are also realized in post-production, i.e., the software bill of materials and CVE databases are continuously monitored to catch newly disclosed vulnerabilities. In this way, the CDT approach enables vulnerability management of the entire lifecycle of a firmware.

Our approach has been implemented in the CDT Platform~\cite{CTL2020}. We applied our approach to about 100 automotive-oriented firmwares, which exist in almost every car nowadays, such as in-vehicle infotainment (IVI) systems, telematics or advanced driver assistance systems (ADAS).

Most of the analyzed firmwares are \emph{rich components}, which are firmwares that include an operating system (OS), several interfaces and many open-source libraries. Furthermore, they usually have an ARM CPU architecture and a Linux, QNX, or Android OS. The other kind of analyzed firmwares are \emph{lower-level microcontrollers}. Such firmwares are usually of smaller size, have no OS nor open-source libraries, and have a Renesas or Tricore CPU architecture.

The main characteristics that influence creation and analysis time of a CDT are the firmware file size and the number of files.
The file size typically ranges from a few megabytes for small, microcontroller based components, up to tens of gigabytes for complex infotainment systems. The number of files typically ranges from one file up to hundreds of thousands of files.
We typically have run times for a complete CDT creation and analysis that take minutes for small components and a few hours for the largest components.

The presented results in the following are from the third quarter of 2020. 
The top 3 most used open-source libraries on the analyzed automotive firmwares were:

\begin{enumerate}
	\item \textbf{Zlib} 1.2.8 (or older), version date: 2013-04-29, latest release: 2017-01-15, age: 3 years
	\item \textbf{Bzip} 1.0.6, version date: 2010-09-06, latest release: 2019-07-13, age: 8 years
	\item \textbf{Expat} 2.2.0, version date: 2016-06-21, latest release: 2019-09-25, age: 3 years
\end{enumerate}

The \textit{version date} indicates when the used library version was released, while the \emph{latest release} indicates the date of the latest version of the library (at the time the results were processed). 
	
\emph{Zlib} and \emph{Bzip} are libraries for data compression, while \emph{Expat} is a library for parsing XML documents.
One important observation is that all these libraries are at least 3 years outdated, \textit{Bzip} is even 8 years old. This is a serious concern, since using outdated libraries expose firmwares to a higher risk of vulnerabilities than using up-to-date libraries.

The average number of used libraries per firmware (rich components) corresponds to about 150 libraries. 
We can see that amongst the top 3, there are the \emph{Zlib} and \emph{Bzip} libraries that are used for data compression. They are very likely to be used in any kind of firmware, i.e., not only in automotive-oriented firmwares. Libraries that have this kind of characteristic and that can be used across different OS's are naturally going to appear in many components, and therefore, are very popular.

With respect to security vulnerabilities, the two most common publicly known vulnerabilities (CVEs) detected in the analysed firmwares were both found in the library \emph{sqlite} and are use-after-free vulnerabilities (CVE-2020-11656 and CVE-2020-13631). The average number of detected CVEs per firmware (rich component) was about 600 vulnerabilities, from which about 20 were of critical severity.

Finally, the average number of discovered unknown vulnerabilities per firmware is about 80, from which 10 are of high severity. Unknown vulnerabilities are classified based on the types defined by the common weakness enumeration (CWE) database. The three most common unknown vulnerabilities identified in the analyzed firmwares are related to memory corruption (CWE-119), buffer overflow (CWE-125) and access control (CWE-338). The latter unknown vulnerability can lead to the non-fulfillment of the aforementioned exemplary requirement (firmwares shall prevent unauthorized access to privacy information), since inadequate access control can facilitate unauthorized access to privacy information.

\section{4. CONCLUSION}

In future work, we plan to extend our approach to automatically generate test cases for the CDT based on common cybersecurity requirements of automotive software. From the software bill of materials, we get a list of libraries used and the deployed versions. We can use this information to generate fuzz test cases in several ways. The first and most straightforward way would be to use specific protocol fuzzers for libraries that use known protocols. Second, for the open-source libraries, we can retrieve the interfaces and potentially mine input grammars from the repositories directly. Third, for commercial libraries, the further control and data flow analyses on the binaries can be used for fuzzing. Fourth, we could build a database of popular libraries, their interfaces and interesting attach surfaces to fuzz corresponding tests.

\bibliographystyle{ieeetr}
\bibliography{cdt}

\begin{thebibliography}{10}

\bibitem{Fernandez2017}
D.~M. Fern{\'a}ndez, S.~Wagner, M.~Kalinowski, M.~Felderer, P.~Mafra,
  A.~Vetr{\`o}, T.~Conte, M.-T. Christiansson, D.~Greer, C.~Lassenius, {\em
  et~al.}, ``Naming the pain in requirements engineering,'' {\em Empirical
  software engineering}, vol.~22, no.~5, pp.~2298--2338, 2017.

\bibitem{UniNations2020}
{United Nations}, ``{UNECE WP.29 regulation},'' 2020.

\bibitem{ISO-21434-2020}
{International Organization for Standardization.}, ``{ISO/SAE DIS 21434 Road
  vehicles - Cybersecurity engineering},'' 2020.

\bibitem{Cruzes2017}
D.~S. Cruzes, M.~Felderer, T.~D. Oyetoyan, M.~Gander, and I.~Pekaric, ``How is
  security testing done in agile teams? a cross-case analysis of four software
  teams,'' in {\em International Conference on Agile Software Development},
  pp.~201--216, Springer, Cham, 2017.

\bibitem{Ciravegna2013}
L.~Ciravegna, P.~Romano, and A.~Pilkington, ``{Outsourcing practices in
  automotive supply networks: an exploratory study of full service vehicle
  suppliers},'' {\em International Journal of Production Research}, vol.~51,
  no.~8, pp.~2478--2490, 2013.

\bibitem{Liu2012}
B.~Liu, L.~Shi, Z.~Cai, and M.~Li, ``{Software Vulnerability Discovery
  Techniques: A Survey},'' in {\em 2012 Fourth International Conference on
  Multimedia Information Networking and Security}, pp.~152--156, 2012.

\bibitem{Boschert2016}
S.~Boschert and R.~Rosen, {\em {Digital Twin--The Simulation Aspect}},
  pp.~59--74.
\newblock Cham: Springer International Publishing, 2016.

\bibitem{CTL2020}
{Cybellum Technologies Ltd}, ``{Introduction to Cybellum Security Suite –
  Risk Assessment for Automotive Software},'' 2020.

\bibitem{Chapman2011}
P.~Chapman and D.~Evans, ``{Automated black-box detection of side-channel
  vulnerabilities in web applications},'' in {\em Proceedings of the 18th ACM
  conference on Computer and communications security}, pp.~263--274, 2011.

\bibitem{Djuric2013}
Z.~Djuric, ``{A black-box testing tool for detecting SQL injection
  vulnerabilities},'' in {\em 2013 Second international conference on
  informatics \& applications (ICIA)}, pp.~216--221, IEEE, 2013.

\bibitem{Grieves2017}
M.~Grieves and J.~Vickers, ``Digital twin: Mitigating unpredictable,
  undesirable emergent behavior in complex systems,'' in {\em Transdisciplinary
  perspectives on complex systems}, pp.~85--113, Springer, 2017.

\bibitem{Glaessgen2012}
E.~Glaessgen and D.~Stargel, ``{The digital twin paradigm for future NASA and
  US Air Force vehicles},'' in {\em 53rd AIAA/ASME/ASCE/AHS/ASC structures,
  structural dynamics and materials conference 20th AIAA/ASME/AHS adaptive
  structures conference 14th AIAA}, p.~1818, 2012.

\bibitem{Hadar2020}
E.~Hadar, D.~Kravchenko, and A.~Basovskiy, ``{Cyber Digital Twin Simulator for
  Automatic Gathering and Prioritization of Security Controls’
  Requirements},'' in {\em 2020 IEEE 28th International Requirements
  Engineering Conference (RE)}, pp.~250--259, IEEE, 2020.

\end{thebibliography}

\begin{wrapfigure}{l}{0.2\textwidth}
\includegraphics[width=0.2\textwidth]{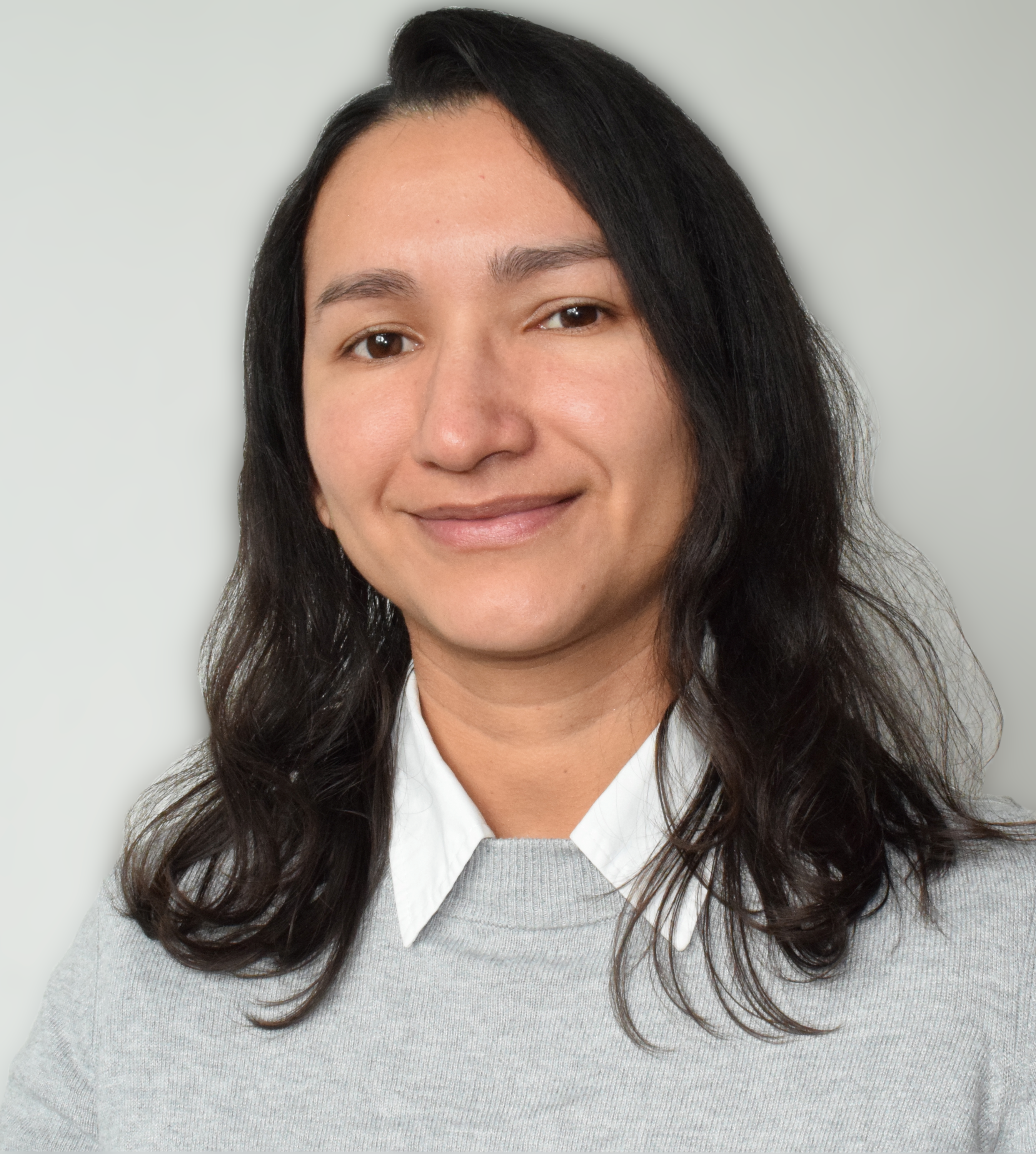}
\end{wrapfigure}
\begin{IEEEbiography}{Ana Cristina Franco da Silva}{\,} is a postdoc at the Institute of Software Engineering at the University of Stuttgart. Her research interests include testing in the IoT, data processing in IoT environments, and software quality. She holds a bachelor degree in computing engineering, a master degree in software engineering, and defended her Ph.D. in the field of data processing in the IoT at the University of Stuttgart. Contact her at ana-cristina.franco-da-silva@iste.uni-stuttgart.de.\\\\
\end{IEEEbiography}

\begin{wrapfigure}{l}{0.2\textwidth}
\includegraphics[width=0.2\textwidth]{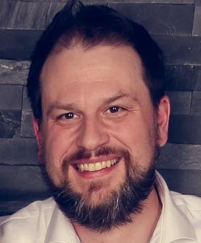}
\end{wrapfigure}
\begin{IEEEbiography}{Stefan Wagner}{\,} is a full professor of empirical software engineering and managing director of the Institute of Software Engineering at the University of Stuttgart. His research interests are requirements engineering, software quality, safety and security engineering, agile and continuous software development. He studied computer science in Augsburg and Edinburgh and received a doctoral degree from the Technical University of Munich. Contact him at stefan.wagner@iste.uni-stuttgart.de.\\\\\\
\end{IEEEbiography}

\begin{wrapfigure}{l}{0.2\textwidth}
\includegraphics[width=0.2\textwidth]{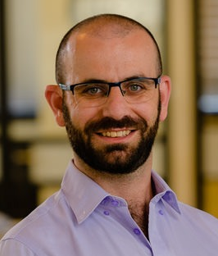}
\end{wrapfigure}
\begin{IEEEbiography}{Eddie Lazebnik}{\,} is Global Partners Director at Cybellum, leading the company’s strategic partnerships activity. Eddie has 15 years of experience in cybersecurity, both in the private and public sectors including the Israeli government and military organizations. Eddie holds a B.Sc. in Electrical Engineering and Electronics, B.Sc. in Physics and an MBA, all from Tel Aviv University. Contact him at eddie@cybellum.com.
\end{IEEEbiography}

\begin{wrapfigure}{l}{0.2\textwidth}
\includegraphics[width=0.2\textwidth]{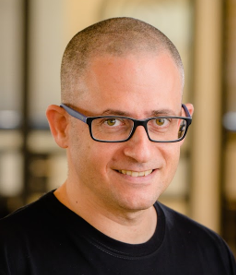}
\end{wrapfigure}
\begin{IEEEbiography}{Eyal Traitel}{\,} is VP Product Management at Cybellum. In his role, he is managing the product planning, from requirements to design and UX/UI, working closely with customers to understand their needs in product security, and following regulatory requirements, analyzing standards such as WP.29, ISO 21434 and others. Eyal has over 24 years of global experience in the enterprise software industry, in early stage startups and the largest corporations. Eyal has co-authored books for O’Reilly and IBM Redbooks. Contact him at eyal@cybellum.com.
\end{IEEEbiography}

\end{document}